\documentclass[a4paper,11pt]{article}
\pdfoutput=1

\usepackage{jcappub}
\usepackage{graphicx}
\graphicspath{ {paper1_method/} }
\usepackage{dcolumn}
\usepackage{amssymb,amsmath,bm}
\usepackage{mathtools}
\usepackage{color}
\usepackage[dvipsnames]{xcolor}
\usepackage{xfrac}
\usepackage{aas_macros}
\usepackage{mathrsfs}
\usepackage{subcaption}
\usepackage{rotating}
\usepackage{chngcntr}
\usepackage[T1]{fontenc} % if needed
\usepackage{natbib}
\usepackage{lipsum}
\newcommand{\nv}{\hat{\bm n}}
\newcommand{\xv}{{\bm x}}
\newcommand{\kv}{{\bm k}}
\newcommand{\Ninj}{N_\mathrm{inj}}
\newcommand{\Einj}{E_\mathrm{inj}}
\newcommand{\tinj}{t_\mathrm{inj}}

\newcommand{\deq}{\coloneqq}

\newcommand{\Ecut}{E_\mathrm{cut}}
\newcommand{\cN}{\mathcal{N}}
\newcommand{\cS}{\mathcal{S}}
\newcommand{\sN}{\mathrm{\textsc n}}
\newcommand{\sS}{\mathrm{\textsc s}}
\newcommand{\de}{\mathrm{d}}

\title{Detecting ultra-high energy cosmic ray anisotropies through harmonic cross-correlations}
\author[a]{Federico R.~Urban,}
\author[b,c]{Stefano Camera,}
\author[d]{and David Alonso}

\affiliation[a]{CEICO, Institute of Physics of the Czech Academy of Sciences,\\Na Slovance 1999/2, 182 21 Prague, Czech Republic}
\affiliation[b]{Dipartimento di Fisica, Universit\`a degli Studi di Torino,\\Via P.\ Giuria 1, 10125 Torino, Italy}
\affiliation[c]{INFN -- Istituto Nazionale di Fisica Nucleare, Sezione di Torino,\\Via P.\ Giuria 1, 10125 Torino, Italy}
\affiliation[d]{Department of Physics, University of Oxford,\\Denys Wilkinson Building, Keble Road, Oxford OX1 3RH, UK}

\emailAdd{federico.urban@fzu.cz}

\abstract{We propose an observable for ultra-high energy cosmic ray (UHECR) physics: the \emph{harmonic-space} cross-correlation power spectrum between the arrival directions of UHECRs and the large-scale cosmic structure mapped by galaxies. This cross-correlation has not yet been considered in the literature, and it permits a direct theoretical modelling of the main astrophysical components. We describe the expected form of the cross-correlation and show how, if the distribution of UHECR sources trace the large-scale cosmic structure, it could be easier to detect with current data than the UHECR auto-correlation. Moreover, the cross-correlation is more sensitive to UHECR anisotropies on smaller angular scales, more robust to systematic uncertainties, and it could be used to determine the redshift distribution of UHECR sources, making it a valuable tool in determining their origins and properties.}

\begin{document}
\maketitle
\flushbottom

% --  --  --  --  --  --  --  --  --  --  --  --  --  --  --  --  --  --  --  --  --  --  --  --  --  -- -
\section{Introduction}\label{sec:intro}
% --  --  --  --  --  --  --  --  --  --  --  --  --  --  --  --  --  --  --  --  --  --  --  --  --  -- -

Ultra-high energy cosmic rays (UHECRs), impacting the atmosphere of the Earth with energies in excess of \(1\,\mathrm{EeV}\) (\(10^{18}\,\mathrm{eV}\)), have remained a mystery since their discovery 59 years ago \cite{Linsley:1961kt,AlvesBatista:2019tlv}.  We do not know what they are: observational data can not yet fully distinguish between several variants of pure and mixed primary compositions \cite{Castellina:2019huz,Bergman:2019aaa}.  We do not know where they come from: the astrophysical sources that generate and accelerate UHECRs have not been identified yet; the type of acceleration mechanism that is responsible for their formidable energies has not been discovered, either \cite{Kotera:2011cp}.

What we do know is that the highest energy rays are most likely extra-galactic.  First, if UHECRs were produced within the Galaxy, their arrival directions in the sky would be very different from what we observe \cite{Tinyakov:2015qfz,Abbasi:2016kgr,Aab:2017tyv}.  Second, barring a cosmic conspiracy that puts an end to the injection spectrum at that very energy, UHECR interactions with cosmological background photons produce a sharp cutoff (the Greisen-Zatsepin-Kuzmin limit) in the spectrum corresponding to \(\sim60\,\mathrm{EeV}\) \cite{Greisen:1966jv,Zatsepin:1966jv}, and a cutoff is indeed observed in the data \cite{Abbasi:2007sv,Abraham:2008ru}.

If the sources of UHECRs are extra-galactic, they most probably correlate with the large-scale distribution of matter (large-scale structure, or LSS). The interactions with the background cold photons limit UHECR propagation to roughly a few hundreds of Mpc (for a review, see Ref.~\cite{Kotera:2011cp}).  Therefore, the UHECR flux distribution in the sky should be to some extent anisotropic, since below \(100\,\mathrm{Mpc}\), roughly comparable with the scale of homogeneity expected in the standard cosmological model, LSS are anisotropic \cite{Pan:2000yg,Scrimgeour:2012wt,Alonso:2014xca}.

How the anisotropy of UHECR sources manifests itself in the observed flux on Earth then depends on the original anisotropy of the sources, the UHECR chemical composition, and the properties of intervening magnetic fields -- Galactic (GMF) and extra-galactic (xGMF) -- that deflect UHECRs and distort the original anisotropic patterns. Chemical composition and magnetic fields are degenerate when it comes to UHECRs deflections, since the latter depends on \(ZB/E\), where \(Z\) is the atomic number, \(B\) the strength of the magnetic field, and \(E\) is the UHECR energy: doubling the field strength is equivalent to doubling the charge (or halving the energy). Chemical composition instead is the only factor that determines the UHECR propagation length at a given energy: different nuclei come from different portions of the Universe and carry different anisotropic imprints, but the relationship between the two is non-monotonic and non-trivial (see, e.g., \cite{dOrfeuil:2014qgw,diMatteo:2017dtg}).

To a large extent, the statistics of the anisotropies in the distribution of UHECRs can be characterized by the UHECR angular auto-correlation (AC), which, in harmonic space, takes the form of the angular power spectrum coefficients \(C_\ell\). Here, the \(\ell\)-th multipole quantifies the variance of the anisotropies on angular scales \(\theta\sim\pi/\ell\) \cite{Sommers:2000us,Tinyakov:2014fwa} (see Appendix \ref{app:pk_cl} for further details).  To date, the number of UHECRs collected at the highest energies is low -- of the order of a hundred above the cutoff \cite{AlvesBatista:2019tlv}.  Because of this, the UHECR flux is dominated by Poisson statistics: the AC is mostly determined by shot noise, making the underlying correlation with the LSS very hard to detect. Indeed, the indications for anisotropy in the data are tenuous: save for a low-energy dipole \cite{Aab:2017tyv} and a high-energy hot-spot \cite{Abbasi:2014lda}, the angular distribution of UHECR arrival directions appears to be nearly isotropic \cite{diMatteo:2020dlo}.  Moreover, no anisotropies have been detected at small scales \(\ell\gtrsim10\) \cite{Deligny:icrc2015,diMatteo:2018vmr}, although there are hints at intermediate scales \cite{Caccianiga:2020njq}.

In this work, we quantify the possibility of detecting the anisotropy in the UHECR flux through the harmonic-space power spectrum of the cross-correlation (XC) between UHECR counts and the distribution of galaxies.  Such XC technique was previously proposed to study the anisotropy of the \(\gamma\)-ray sky by Refs.~\cite{Camera:2012cj,Fornengo:2013rga,Pinetti:2019ztr}, and proved successful for several tracers of the LSS \cite{Fornengo:2014cya,Cuoco:2015rfa,Branchini:2016glc,Ammazzalorso:2019wyr}. A search for a XC between UHECRs and high-energy photons was performed in \cite{Alvarez:2016otl}.  If UHECR sources statistically trace the LSS, then the positions of these sources, and the arrival directions of UHECRs (if not strongly affected by intervening magnetic fields) should have a non-zero correlation with a galaxy sample up to a given distance.  Therefore, the detection or non-detection of the XC signal with galaxies at different redshifts would allow us to test whether UHECR sources are distributed according to the LSS, and to quantify the extent to which the UHECR transfer function, determined by energy losses and intervening magnetic fields, does not depend on direction.

There are at least three features that differentiate the XC from other methods (see for instance \cite{Koers:2008ba} and references therein).  First, systematic uncertainties of different `messengers', or observables, should not cross-correlate, and, under some conditions, statistical noise should also not strongly cross-correlate.  This is because different experiments are different machines exploiting different physical effects.  However, within a single data set, for instance the set of arrival directions of UHECRs, the AC of the noise and systematic errors for that set are certainly non-zero, and contribute to hiding any underlying `true' signal.  Examples of these systematics for UHECRs would be perturbations in the arrival directions due to deflection by the GMF\footnote{Magnetic deflections can also affect the size of a data set in the case of partial sky coverage (where UHECR events could migrate to/from the observed patch of the sky).}, or spatial fluctuations in the energy calibration giving rise to leakage from different energy bins\footnote{Overall energy miscalibration would still need to be modelled and accounted for in the theory predictions.}.  Thus, in this sense the XC is an experimentally cleaner observable.

Secondly, in the limit where the UHECR sources are numerous, but UHECR detections themselves are not, we can assume that we observe at most one UHECR per source (as seems to be the case given to the lack of obvious UHECR multiplets \cite{Abreu:2011md,Aab:2019ogu}). The much higher number of galaxies leads to a significant improvement in the signal-to-noise ratio of this cross-correlation (see the discussion in section~\ref{sec:results}).  This effectively allows us to probe the anisotropies on smaller scales through the XC than the AC, underlying the importance of using both observables.

There are several reasons why those smaller scales (\(\ell>10\)) are interesting.  First of all, the experimental angular resolution of UHECR events is around \(1^\circ\), which corresponds to \(\ell\sim200\): from an experimental perspective we are not fully taking advantage of the data we already have.  Furthermore, small-scale power in the LSS angular distribution is comparable to that at large scales: if UHECRs bear the imprint of the LSS this small-scale power is not completely suppressed by the GMF, especially once the structured component of the GMF is taken into account \cite{Dundovic:2017vsz}; moreover, the sub-structures of the GMF themselves imprint small-scale anisotropies in PeV cosmic rays \cite{Giacinti:2011mz} and it is possible that such structures can be present at higher energies.  Lastly, small-scale anisotropies can be separately detected in different regions of the sky, allowing us to probe, for example, different GMF structures independently.

Third, while most analyses have looked at the real-space correlation between UHECRs and the large-scale structure \cite{Kashti:2008bw,Oikonomou:2012ef,Abreu:2010ab,PierreAuger:2014yba,Takami:2008ri}), we will express our results here in terms of harmonic-space power spectra.  These are common observables in cosmological studies, based on a natural decomposition of the celestial sphere.  They also allow for a straightforward visualization of the main components of the astrophysical model (radial kernels, details of the galaxy-matter connection), which is one of the main novel aspects of this work.

In this paper, we will introduce a formalism to model the AC and XC, and apply it to a vanilla proton-only model for UHECR injection in order to quantify the differences between the two observables and the detectability of the anisotropies on different scales with existing experimental facilities.  We defer the more detailed discussion of the dependence of the XC on UHECR injection and source properties, a realistic treatment of the UHECR experimental setup, such as non-uniform sky coverage, as well as a full treatment of the effects of the GMF and xGMF on the signal, to upcoming work.

This paper is organized as follows.  In Section~\ref{sec:model} we introduce the formalism to describe the UHECR flux, the distribution of galaxies, and the AC and XC.  We apply this formalism to a hypothetical UHECR dataset in Section~\ref{sec:results}, where we obtain and compare the AC and the XC.  We summarize our findings and conclude with an outlook for future work in Section~\ref{sec:conclusions}.  Appendix \ref{app:pk_cl} collects useful standard formul\ae~pertaining to angular power spectra.

% --  --  --  --  --  --  --  --  --  --  --  --  --  --  --  --  --  --  --  --  --  --  --  --  --  -- -
\section{Theoretical model}\label{sec:model}
% --  --  --  --  --  --  --  --  --  --  --  --  --  --  --  --  --  --  --  --  --  --  --  --  --  -- -

% --  --  --  --  --  --  --  --  --  --  --  --  --  --  --  --  --  --  --  --  --  --  --  --  --  -- -
\subsection{UHECR flux}\label{ssec:model.flux}

Let \(\mathcal{E}(\Einj)\) be the (angle-integrated, isotropic) emissivity\footnote{Note that our definition of emissivity differs from the one used in, e.g.\ radio astronomy, which quantifies the \emph{energy} (instead of \emph{number}) emitted per unit time, volume, and solid angle.} of cosmic rays (CRs) for a given galaxy (number of CRs of energy \(\Einj\) emitted per unit energy, per unit time):
\begin{align}
	\mathcal{E}[\Einj]\deq\frac{\de\Ninj}{\de\Einj\,\de\tinj} \,.
\end{align}
The subscript `inj' (injection) here indicates quantities evaluated in the rest frame of the emitting source.  Due to the expansion of the Universe and to interactions between CRs and cosmic background light, the injected energy of a CR, whose energy at detection is \(E\), is given by \(\Einj(E,z)\) with \(z\) the redshift of the source. In the absence of scattering processes the energy losses are adiabatic \(\Einj = (1+z)E\). The differential emissivity (i.e., per unit solid angle) is \(\epsilon\deq\mathcal{E}/4\pi\) assuming isotropic emission.  We will parameterise the emissivity as a power-law of energy:
\begin{align}\label{eq:plaw}
	\mathcal{E}[\Einj]\propto\Einj^{-\gamma} \,.
\end{align}
Energies will always be expressed in EeV for convenience.
 
The quantity measured on Earth is the observed number of events per unit time, energy interval, detector area, solid angle on the sky and (assuming source redshifts can be measured), redshift interval. We can relate this number to the emissivity through
\begin{equation}
	\frac{\de N}{\de E\,\de t\,\de A\,\de \Omega\,\de z}=\frac{ n_{\rm s,c}\,\mathcal{E}(\Einj)}{4\pi\,(1+z)\,H(z)}\frac{\de \Einj}{\de E} \,,
\end{equation}
where \(H(z)\) is the Hubble parameter, \(n_{\rm s,c}\) is the volumetric number density of CR sources, we have set \(c=1\), and we have ignored subdominant light-cone and relativistic effects \citep{Challinor:2011bk,Bonvin:2011bg}.

We will be interested in the number of UHECRs detected above a given energy threshold \(\Ecut\) (defined in the observer's frame) and integrated over source redshifts, from the direction \(\nv\):
\begin{align}\nonumber
	\Phi(\Ecut,\nv)&\deq \int_0^\infty \de z\;\int_{\Ecut}^\infty \de E\;\frac{\de N}{\de E\,\de t\,\de A\,\de \Omega\,\de z}\\
	&=\int \frac{\de z}{(1+z)H(z)}\;\frac{n_{\rm s,c}(z,\chi\nv)}{4\pi}\,\int_{\Ecut}^\infty \de E\;\frac{\de \Einj}{\de E}\,\mathcal{E}(\Einj) \,,
\end{align}
where \(\chi(z)\) is the radial comoving distance.

We can write the number density of sources as \(n_{\rm s,c}(z,\chi\nv)=\bar{n}_{\rm s,c}(z)\,[1+\delta_{\rm s}(z,\chi\nv)]\), where \(\delta_{\rm s}\) is the galaxy overdensity. Assuming a non-evolving galaxy population, namely \(\bar{n}_{\rm s,c}(z) = \bar{n}_{\rm s,c}(0)\), and a power-law UHECR spectrum (as in Eq.~\ref{eq:plaw}) we obtain:
\begin{align}%\nonumber
	\Phi(\Ecut,\nv)&\propto\frac{\bar{n}_{\rm s,c}}{4\pi}\int \frac{\de \chi}{(1+z)}\;\frac{\Einj^{1-\gamma}(\Ecut,z)}{1-\gamma}\,\left[1+\delta_{\rm s}(z,\chi\nv)\right] \,.
\end{align}

% --  --  --  --  --  --  --  --  --  --  --  --  --  --  --  --  --  --  --  --  --  --  --  --  --  -- -
\subsubsection{Attenuation}\label{sssec:model.flux.attenuation}

The \emph{attenuation} factor \(\alpha(\Ecut,z;\gamma,Z)\) is defined as the number of events reaching the Earth with \(E>\Ecut\) divided by the number of events which would have reached the Earth if there were no energy losses at a given distance:
\begin{equation}
  \alpha(z,\Ecut;\gamma,Z)\deq\frac{\Einj^{1-\gamma}(\Ecut,z)}{\Ecut^{1-\gamma}}\,.
\end{equation}
The attenuation \(\alpha\) is a function of the energy cut and redshift, as well as the injection spectral slope and chemical composition. In terms of \(\alpha\), the direction-dependent integral flux is
\begin{align}
  \Phi(\Ecut,\nv)\propto\frac{\bar{n}_{\rm s,c}\Ecut^{1-\gamma}}{4\pi\,(1-\gamma)}\int \de\chi\;\frac{\alpha(z,\Ecut)}{(1+z)}\,\left[1+\delta_{\rm s}(z,\chi\nv)\right].
\end{align}

\begin{figure}
\centering
  \includegraphics[width=0.75\textwidth]{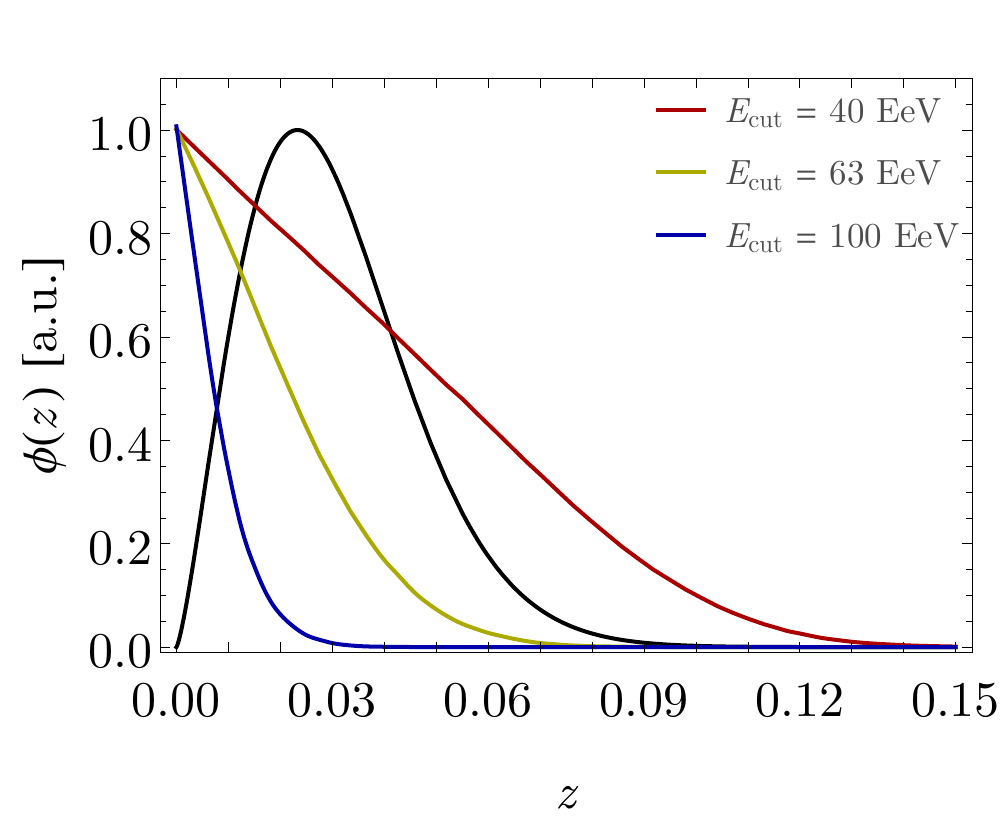}
   \caption{Radial kernels for the two observables under consideration. The solid black line shows the approximate redshift distribution of galaxies in the 2MRS sample using the fit found by \cite{Ando:2017wff}. The red, yellow, and blue lines show the radial kernel for the UHECR flux (Eq.~(\ref{eq:window_cr})) for the three energy thresholds studied here (\(40\,\mathrm{EeV}\), \(63\,\mathrm{EeV}\), and \(100\,\mathrm{EeV}\) respectively).}
   \label{fig:kernels}
\end{figure}

In this paper, to introduce the formalism, we choose to work with a toy proton-only model with injection slope \(\gamma=2.6\) as in model (4) of \cite{dOrfeuil:2014qgw}, or model (i) of \cite{diMatteo:2017dtg}.  In order to obtain the attenuation factor for our injection model we have followed \(10^6\) events with \emph{SimProp}~v2r4 \cite{Aloisio:2017iyh} with energies above \(E=10\,\mathrm{EeV}\) (with an upper cutoff of \(E=10^{5}\,\mathrm{EeV}\)), for redshifts up to \(z=0.3\), and counted the number of events reaching the Earth with \(E>\Ecut\) for different values of \(\Ecut\).  With \emph{SimProp} we have accounted for all energy losses, adiabatic and interactions with cosmic microwave background (CMB) photons and extra-galactic background photons according to the model \cite{Stecker:2005qs}. The UHECR radial kernels, defined in the next section, obtained from the attenuation factor \(\alpha\) for different energies are shown in Fig.~\ref{fig:kernels}.

Note that we assume that cosmic ray energy losses are to first order isotropic, that is, we ignore angular anisotropies in in the CMB and extra-galactic background light, which are completely negligible for our analysis.  Moreover, for simplicity here we work with full-sky uniform coverage, but the analysis can be readily generalized to non-uniform and partial sky coverage.

% --  --  --  --  --  --  --  --  --  --  --  --  --  --  --  --  --  --  --  --  --  --  --  --  --  -- -
\subsubsection{Anisotropies}\label{sssec:model.flux.anisotropies}

We will define the anisotropy in the UHECRs distribution as the over-density of rays detected as a function of sky position \(\nv\) as
\begin{equation}
	\Delta_{\rm CR}(\nv,E_{\rm cut}) \deq \frac{\Phi(\nv,E_{\rm cut})-\bar{\Phi}(E_{\rm cut})}{\bar{\Phi}(E_{\rm cut})} \,,
\end{equation}
where \(\bar{\Phi}(\Ecut)\) is the sky-averaged UHECR flux. From the results in the previous section, this quantity is related to the three-dimensional overdensity of UHECR sources \(\delta_{\rm s}(z,\chi\nv)\) through
\begin{equation}\label{eq:delta_cr}
	\Delta_{\rm CR}(\nv,\Ecut)=\int \de\chi\;\phi_{\rm CR}(\chi)\,\delta_{\rm s}(z,\chi\nv) \,,
\end{equation}
where the UHECR radial kernel (or window function) is
\begin{equation}\label{eq:window_cr}
	\phi_{\rm CR}(z)\deq\left[\int \de\tilde z\;\frac{\alpha(\tilde z)}{H(\tilde z)(1+\tilde z)}\right]^{-1}\frac{\alpha(z)}{(1+z)} \,.
\end{equation}
Figure \ref{fig:kernels} shows the radial kernels for UHECRs with energy thresholds \(\Ecut=40,\,63\), and \(100\,\mathrm{EeV}\); as expected the lower the energy the farther UHECRs propagate.

% --  --  --  --  --  --  --  --  --  --  --  --  --  --  --  --  --  --  --  --  --  --  --  --  --  -- -
\subsection{Galaxies}\label{ssec:model.gals}

We will consider the AC of the UHECR anisotropy, Eq.~\eqref{eq:delta_cr}, and its XC with the galaxy number count fluctuations.  In particular, we will work with the projected overdensity of sources for a given galaxy sample,
\begin{equation}
    \Delta_{\rm g}(\nv)\deq\frac{N_{\rm g}(\nv)-\bar{N}_{\rm g}}{\bar{N}_{\rm g}},
\end{equation}
where \(N_{\rm g}(\nv)\) is the number of galaxies in a given direction \(\nv\), and \(\bar{N}_{\rm g}\) its average over the celestial sphere.  This is related to the three-dimensional galaxy overdensity \(\delta_{\rm g}(z,\chi\,\nv)\) via
\begin{equation}\label{eq:delta_g}
	\Delta_{\rm g}(\nv)=\int \de\chi\;\phi_{\rm g}(\chi)\,\delta_{\rm g}(z,\chi\,\nv) \,,
\end{equation}
where \(\phi_{\rm g}(\chi)\) is the weighted distribution of galaxy distances. In general, we will assume that we have redshift information for all galaxies in the catalog, and that we can use that information to apply a distance-dependent weight \(w(\chi)\). In that case, the galaxy overdensity kernel \(\phi_{\rm g}(\chi)\) is given by
\begin{align}\label{eq:window_g}
	\phi_{\rm g}(\chi)\deq \left[\int \de\tilde\chi\; \tilde\chi^2\,w(\tilde\chi)\,\bar{n}_{\rm g,c}(\tilde\chi)\right]^{-1}\,\chi^2\,w(\chi)\,\bar{n}_{\rm g,c}(\chi) \,,
\end{align}
where \(\bar{n}_{\rm g,c}\) is the comoving number density of galaxies in the sample.

If no weights are applied -- namely, \(w(\chi)=1\) -- then
\begin{equation}\label{eq:n3d_n2d}
	\int \de\chi\;\chi^2\,w(\chi)\,\bar{n}_{\rm g,c}(\chi) = \bar{N}_{\Omega,{\rm g}} \,,
\end{equation}
where \(\bar{N}_{\Omega,{\rm g}}\) is the angular number density of galaxies (i.e., number of galaxies per steradian).

Figure \ref{fig:kernels} shows the radial kernel for a low-redshift galaxy survey, modelled after the 2MASS Redshift Survey (2MRS) \cite{2012ApJS..199...26H}. This constitutes one of the most complete full-sky spectroscopic low-redshift surveys, and we will use it as our fiducial galaxy sample in this paper.  In this work we consider full-sky data sets for simplicity, but the generalization of our results for an incomplete sky coverage is straightforward.  In the case of a realistic setup based on 2MRS, a sky coverage around 70\% will only degrade the signal by a factor of \(\sqrt{0.7}\simeq0.86\).

% --  --  --  --  --  --  --  --  --  --  --  --  --  --  --  --  --  --  --  --  --  --  --  --  --  -- -
\subsection{Power spectra}\label{ssec.model.cls}

We are interested in detecting the intrinsic anisotropies in the distribution of UHECRs by considering the different two-point functions built from \(\Delta_{\rm CR}\) and \(\Delta_{\rm g}\). A given observation of any of these fields will consist of both signal \(\sS\) and noise \(\sN\): \(\Delta_a=\sS_a+\sN_a\) (where \(a,\,b\,\in\{{\rm CR},g\}\)). Assuming signal and noise to be uncorrelated, the corresponding power spectra can be split into both components, namely
\begin{align}
    C_\ell\deq\cS_\ell+\cN_\ell \,,
\end{align}
where \(\cS_\ell\) and \(\cN_\ell\) are the power spectra of \(\sS\) and \(\sN\) respectively. In our case, the signal is the intrinsic clustering of both UHECRs and galaxies due to the underlying large-scale structure, while the noise is sourced by the discrete nature of both tracers as Poisson noise. A brief review of the mathematics behind angular power spectra is given in Appendix \ref{app:pk_cl}.

% --  --  --  --  --  --  --  --  --  --  --  --  --  --  --  --  --  --  --  --  --  --  --  --  --  -- -
\subsubsection{Signal power spectra}\label{sssec:model.cls.sls}

The angular power spectrum \(\cS_\ell^{ab}\) between two projected quantities \(\Delta_a\) and \(\Delta_b\) is related to their three-dimensional power spectrum \(P_{ab}(z,k)\) by
\begin{equation}\label{eq:cl_limber}
	\cS^{ab}_\ell=\int \frac{\de\chi}{\chi^2}\;\phi_a(\chi)\,\phi_b(\chi)\,P_{ab}\left[z(\chi),k=\frac{\ell+1/2}{\chi}\right] \,,
\end{equation}
where \(\phi_a\) and \(\phi_b\) are the radial kernels of both quantities.

The final piece of information needed in order to estimate the expected AC and XC signals is the power spectrum of the three-dimensional overdensities \(\delta_{\rm s}\) and \(\delta_{\rm g}\). In general, the clustering properties of galaxies and UHECR sources will depend on the specifics of the relationship between galaxies and dark matter, and on the astrophysical properties of the UHECR sources. To simplify the discussion, here we will assume that all UHECR sources are also galaxies of the 2MASS sample (i.e.\ \(\delta_{\rm s}=\delta_{\rm g}\)).

At this point, one might be tempted to use a linear bias prescription \citep{Mo:1995cs} to relate the galaxy and matter power spectra. However, as we show in Section \ref{sec:results}, since the UHECR radial kernel peaks at \(z=0\) and covers only low redshifts, the cosmic ray flux auto-correlation probes mostly sub-halo scales for which a non-perturbative description of structure formation is necessary. To achieve this, we use here a halo model prescription \cite{Peacock:2000qk}, based on the halo occupation distribution model used by Ref.~\cite{Ando:2017wff} to describe the 2MRS sample. In this model, the galaxy power spectrum is given by two contributions,
\begin{equation}
 P_{\rm g\,g}(z,k)=P_{\rm g\,g}^{1{\rm h}}(z,k)+P_{\rm g\,g}^{2{\rm h}}(z,k) \,,
\end{equation}
being the so-called 1-halo and 2-halo terms. The former dominates on small scales and describes the distribution of galaxies within the halo, while the latter is governed by the clustering properties of dark matter haloes. The halo occupation distribution is then based on a prescription to assign central and satellite galaxies to haloes of different masses. Although we have summarized this model in Appendix \ref{app:hod}, we refer the reader to \cite{Ando:2017wff} and references therein, for further details about the specifics of the halo occupation distribution model used.

% --  --  --  --  --  --  --  --  --  --  --  --  --  --  --  --  --  --  --  --  --  --  --  --  --  -- -
\subsubsection{Shot noise}\label{sssec:model.cls.nls}
Both projected overdensities, \(\Delta_{\rm CR}\) and \(\Delta_{\rm g}\), are associated to discrete point processes, represented by the angular positions of the UHECRs and the galaxies in each sample. In that case, even in the absence of intrinsic correlations between the different fields, their power spectra receive a non-zero white contribution, given by
\begin{equation}
	\cN^{ab}_\ell=\frac{\bar{N}_{\Omega,a\bigwedge b}}{\bar{N}_{\Omega,a}\,\bar{N}_{\Omega,b}} \,,
\end{equation}
where \(\bar{N}_{\Omega,a}\) (\(\bar{N}_{\Omega,b}\)) is the angular number density of points in sample \(a\) or \(b\), and \(\bar{N}_{\Omega,a\bigwedge b}\) is the angular number density of points shared in common. In our case this would correspond to the number of UHECRs originating from galaxies in the galaxy sample. For simplicity we will assume that the galaxy survey under consideration is sufficiently complete, so that all UHECRs are associated to an observed galaxy. In this case, the shot-noise contributions to the power spectra are
\begin{align}\label{eq:shot_noises}
	\cN^{{\rm CR\,CR}}_\ell&=\left(\bar{N}_{\Omega,{\rm CR}}\right)^{-1} \,, \\
	\cN^{\rm g\,g}_\ell=\cN^{{\rm g\,CR}}_\ell&=\left(\bar{N}_{\Omega,{\rm g}}\right)^{-1} \,.
\end{align}
Since typically \(\bar{N}_{\Omega,{\rm CR}}\ll \bar{N}_{\Omega,{\rm g}}\), then \(\cN^{{\rm g\,CR}}_\ell\ll \cN^{{\rm CR\,CR}}_\ell\), and therefore we will neglect \(\cN^{{\rm g\,CR}}_\ell\) in what follows. We have explicitly checked that indeed the cross-noise can be safely neglected in all our estimates and numerical results.

Note that, when non-flat weights are applied to the galaxy catalog, the resulting noise power spectrum reads
\begin{align}
  \cN^{\rm g\,g}_\ell = \frac{\int \de\chi\;\chi^2\,w^2(\chi)\,\bar{n}_{\rm g,c}(\chi)}{\left[\int \de\chi\; \chi^2\,w(\chi)\,\bar{n}_{\rm g,c}(\chi)\right]^2} \,.\label{eq:noise_gopt}
\end{align}
For \(w(\chi)=1\), Eq.~\eqref{eq:n3d_n2d} holds, and we recover the result in Eq.~\eqref{eq:shot_noises}.

% --  --  --  --  --  --  --  --  --  --  --  --  --  --  --  --  --  --  --  --  --  --  --  --  --  -- -
\subsubsection{Optimal weights}\label{sssec:model.cls.weight}

We can use the results in this section to derive optimal weights \(w(\chi)\) to maximize the signal-to-noise of the galaxy-UHECR cross-correlation.  Let us pixelise the celestial sphere and consider the UHECRs in a given pixel \(p\), \(\Phi_p\), as well as the vector \(N_{p,i}\) containing the number of galaxies along the same pixel in intervals of distance \(\chi_i\). The optimal weights \(w_i\deq w(\chi_i)\) can be found by maximising the likelihood of \(\Phi_p\) given \(N_{p,i}\) \citep{Alonso:2020mva}, and are given by the so-called \textit{Wiener filter}, i.e.\
\begin{equation}
	w_i = \sum_j \textsf{\textbf{Cov}}^{-1}(N_{p,i},N_{p,j})\,\textsf{\textbf{Cov}}(\Phi_p,N_{p,j}) \,.
\end{equation}
Here \(\textsf{\textbf{Cov}}(x,y)\) is the covariance matrix of two vectors \(x\) an \(y\). 

Assuming Poisson statistics, we can use the results from the previous section to show (see Appendix \ref{app:optweight}) that
\begin{equation}
	w(\chi) = \frac{\alpha[z(\chi),E_{\rm cut};\gamma,Z]}{[1+z(\chi)]\chi^2 \bar{n}_{\rm g,c}(\chi)} \,.
\end{equation}
In hindsight, this result is obvious: by inspecting Eqs.~(\ref{eq:window_cr}) and (\ref{eq:window_g}), we see that the optimal weights modify the radial galaxy kernel \(\phi_{\rm g}\) to make it identical to the UHECR kernel \(\phi_{\rm CR}\), thereby building the most likely estimate of the UHECR flux map from the galaxy positions. As we will see, this involves up-weighting galaxies at low redshifts, from where it is more likely that UHECRs that reach the Earth originate, but few galaxies can be found due to volume effects.  Notice that the weights are completely driven by the theoretical model for UHECR propagation, and do not depend on the actual data (and their errors).  We will show how the use of optimal weights can improve the signal-to-noise ratio for the XC in section~\ref{ssec:results.cls}.

% --  --  --  --  --  --  --  --  --  --  --  --  --  --  --  --  --  --  --  --  --  --  --  --  --  -- -
\subsection{Intervening magnetic fields}\label{ssec:model.MFs}

The Milky Way is host to a magnetic field of a few \(\mu\)G \cite{Boulanger:2018zrk}, which is the screen that befogs UHECR sources.  The variety of parametric models of the GMF, which disagree on the GMF functional forms and parameters, particularly on GMF substructures, reflects the complexity of the GMF, and, at the moment, cannot be taken at face value \cite{Boulanger:2018zrk,Unger:2017kfh}.  As a guideline, we can expect the GMF to deflect a UHECR with energy \(E=100\,\mathrm{EeV}\) by a few degrees for the most part of the sky, except for certain directions close to the Galactic plane \cite{dOrfeuil:2014qgw} (see also Ref.~\cite{Pshirkov:2013wka}).  UHECRs will also be affected by any intervening xGMF, whose strength, shape, and filling factors vary by several orders of magnitude for different models and estimates \cite{Subramanian:2015lua}; however, the xGMF is believed to have a subdominant effect on large-scale UHECR propagation \cite{Pshirkov:2015tua}.

Simple prescriptions to account for part of the effects related to the GMF and the xGMF include smearing the map of sources below a certain angular scale, or mixing that map with an isotropic one (similarly to what is done to take into account catalogue incompleteness beyond a certain distance), see for instance \cite{Koers:2008ba}.  Smearing the sources map in our language is as simple as introducing a (Gaussian) beam in the signals as
\begin{align}
    \cS^{\rm g\,g}_\ell \rightarrow \cS^{\rm g\,g}_\ell {\cal B}_\ell^2 \,,~~&~~\cS^{\rm g\,CR}_\ell \rightarrow \cS^{\rm g\,CR}_\ell {\cal B}_\ell \,,
\end{align}
for the AC and XC, respectively, where
\begin{align}
    {\cal B}_\ell&\deq\frac{1}{\pi\theta_\mathrm{smear}^2}\exp\left[-\ell(\ell+1)\theta_\mathrm{smear}^2\right] \,,
\end{align}
and \(\theta_\mathrm{smear}\) is the smearing angle.

However, these solutions tend to be rather artificial and inaccurately destroy potential signal or structures in the spectra we are looking at.  More precisely, we know that the largest effects due to the small-scale GMF are not isotropic, and in fact vary quite considerably across the sky, see, e.g., \cite{Pshirkov:2013wka,diMatteo:2017dtg}.  More precisely, in \cite{Pshirkov:2013wka} it was found that the UHECR deflections are majorated by the function
\begin{align}\label{eq:beam}
    \theta_\mathrm{smear}(b) & \leq\left(\frac{40\,\mathrm{EV}}{E/Z}\right) \, \frac{1^\circ}{\sin^2b+0.15} \,,
\end{align}
where \(b\) is elevation.  Therefore, if we smear the whole sky with the same smearing angle we are not faithfully representing the sky, and, depending on the smearing angle, we either underestimate the deflections in certain regions or overestimate them in other regions, or both.  Moreover, these solutions do not account for the large-scale galactic field, which is the dominant effect and can not be described by a simple smearing.

For these reasons, and in order to best introduce the method, in this first theoretical work we take a pragmatic approach and, keeping in mind all the caveats listed above, we only discuss briefly the effect of a (constant) smearing angle on the AC and XC (see also Appendix~\ref{app:gmf}), whereas we neglect all other effects of intervening magnetic fields.

% --  --  --  --  --  --  --  --  --  --  --  --  --  --  --  --  --  --  --  --  --  --  --  --  --  -- -
 \section{Results}\label{sec:results}
% --  --  --  --  --  --  --  --  --  --  --  --  --  --  --  --  --  --  --  --  --  --  --  --  --  -- -

% --  --  --  --  --  --  --  --  --  --  --  --  --  --  --  --  --  --  --  --  --  --  --  --  --  -- -
\subsection{Signal-to-noise ratio}\label{ssec:results.fisher}

We estimate the signal-to-noise ratio (SNR) of the UHECR anisotropies as the square root of the Fisher matrix element corresponding to an effective amplitude parameter \(A_{\rm CR}\) multiplying the signal component of \(\Delta_{\rm CR}\) with a fiducial value \(A_{\rm CR}=1\) \cite{2009arXiv0906.0664H}, namely
\begin{align}
  {\rm SNR}^2&\deq\sum_{\ell=\ell_{\rm min}}^{\ell_{\rm max}}\left(\frac{\partial\cS_\ell}{\partial A_{\rm CR}}\right)^{\sf T}\textsf{\textbf{Cov}}^{-1}_{\ell\ell^\prime}\frac{\partial\cS_\ell}{\partial A_{\rm CR}},\\
  &=\sum_{\ell=\ell_{\rm min}}^{\ell_{\rm max}}\left({\rm SNR}_\ell\right)^2,
\end{align}
where \({\bm S}_\ell\) is a vector containing the signal contribution to the power spectra under consideration, \(\textsf{\textbf{Cov}}\) is the covariance matrix of those power spectra, and \({\rm SNR}_\ell\) is the SNR of a single \(\ell\) mode. If the fields being correlated are Gaussian (\(\Delta_{\rm CR}\), \(\Delta_{\rm g}\) in our case), the covariance matrix can be estimated using Wick's theorem to be
\begin{equation}
	\textsf{\textbf{Cov}}\left(C^{ab}_\ell,C^{cd}_\ell\right)=\frac{C^{ac}_\ell C^{bd}_\ell+C^{ad}_\ell C^{bc}_\ell}{(2\ell+1)\Delta\ell}\delta_{\ell\ell^\prime} \,,\label{eq:cov}
\end{equation}
with \(\Delta\ell\) the size of the multipole bin.

At this point we can consider three different cases:
\begin{enumerate}
	\item {\bf AC only.} In this case we only have a measurement of the UHECR AC, \(C^{{\rm CR\,CR}}_\ell\). The SNR is given by
	\begin{equation}\label{eq:sn_auto}
		{\rm SNR}^{\rm CR\,CR}=\sqrt{\sum_{\ell=\ell_{\rm min}}^{\ell_{\rm max}}2(2\ell+1)\left(\frac{\cS^{{\rm CR\,CR}}_\ell}{\cS^{{\rm CR\,CR}}_\ell+\cN^{{\rm CR\,CR}}_\ell}\right)^2} \,.
	\end{equation}
	\item {\bf XC only.} In this case we only use the XC, \(C^{{\rm g\,CR}}_\ell\). The SNR is given by
	\begin{equation}\label{eq:sn_cross}
		{\rm SNR}^{{\rm g\,CR}}=\sqrt{\sum_{\ell=\ell_{\rm min}}^{\ell_{\rm max}}(2\ell+1)\frac{(\cS^{{\rm g\,CR}}_\ell)^2}{C^{\rm g\,g}_\ell C^{{\rm CR\,CR}}_\ell+(C^{{\rm g\,CR}}_\ell)^2}} \,.
	\end{equation}
	\item {\bf All data.} We use all available data, i.e.\ a data vector \({\bm S}_\ell=(\cS^{{\rm CR\,CR}}_\ell,\,\cS^{{\rm g\,CR}}_\ell,\,\cS^{\rm g\,g}_\ell)\). Although this is the manifestly optimal scenario, XCs are arguably safer than ACs in terms of systematic errors, and therefore it is interesting to quantify the loss of information if only XCs are used.
\end{enumerate}
Studying these three cases allows us to explore the benefits of using XCs vs ACs, as well as the relative amount of information in each of the different two-point functions. Given the relatively small number of UHECRs currently measured, shot noise in the UHECR flux is the dominant contribution to the uncertainties. Comparing Eqs.~\eqref{eq:sn_auto} and \eqref{eq:sn_cross}, we can see that the SNR scales like \(N_{\rm CR}^{-1}\) and \(N_{\rm CR}^{-1/2}\) for cases 1 and 2 respectively, highlighting the potential of XCs to achieve a detection.

% --  --  --  --  --  --  --  --  --  --  --  --  --  --  --  --  --  --  --  --  --  --  --  --  --  -- -
\subsection{Power spectra and signal-to-noise}\label{ssec:results.cls}

The energy \(\Ecut\) at which we choose to cut the UHECR integral spectrum determines the UHECR propagation horizon, which in turns determines the strength of the anisotropy.  Moreover, the choice of \(\Ecut\), for a given UHECR spectrum, also determines the number of UHECR events we have to sample the anisotropic angular distribution.  We expect a trade-off between the two.  At low energies the UHECR sample contains many more events than at high energies because the UHECR spectrum is very steep (soft/red); however, for the range of energies we are interested in, the galaxy sample is much larger, so this does not have as strong an effect for the XC as it does for the traditional AC (whose noise is determined by the number of UHECR events).  Moreover, at low energies UHECRs propagate further, and the larger line-of-sight averaging can dilute the expected anisotropy.  Lastly, the effects of intervening magnetic fields are stronger---this is expected to have a significant impact on the anisotropies, albeit less so for the XC compared to the AC thanks to its stability against systematics.  At high energies the UHECR horizon is smaller, UHECRs undergo smaller deflections, and the anisotropy should be more pronounced, but the number of events drops dramatically.

In order to determine at which energy we have the best chances of detecting the XC we chose to work with three energy cuts at: \(\Ecut=10^{19.6}\,\mathrm{eV}\simeq40\,\mathrm{EeV}\), \(\Ecut=10^{19.8}\,\mathrm{eV}\simeq63\,\mathrm{EeV}\), and \(\Ecut=10^{20}\,\mathrm{eV}=100\,\mathrm{EeV}\).  In a realistic scenario, based on data currently available~\cite{AlvesBatista:2019tlv}, we can expect to have about \(N_{\rm CR}=1000\),  \(N_{\rm CR}=200\), and \(N_{\rm CR}=30\) over the full sky, for the three energy cuts defined above, respectively.  While this does not fully reflect a realistic situation, mostly because of magnetic deflections which we do not take into account, and because current experimental facilities are limited in their field of view, our results nonetheless present a fair comparison between the two measures (AC and XC).  This is owing to the fact that all the salient information regarding UHECR data sets is represented in our estimates, namely the energy cut and with it all the UHECR energy losses, the available or expected number of events at that energy, and the angular resolution representative of what current experiments can do.

\begin{figure}
\centering
	\includegraphics[width=\textwidth]{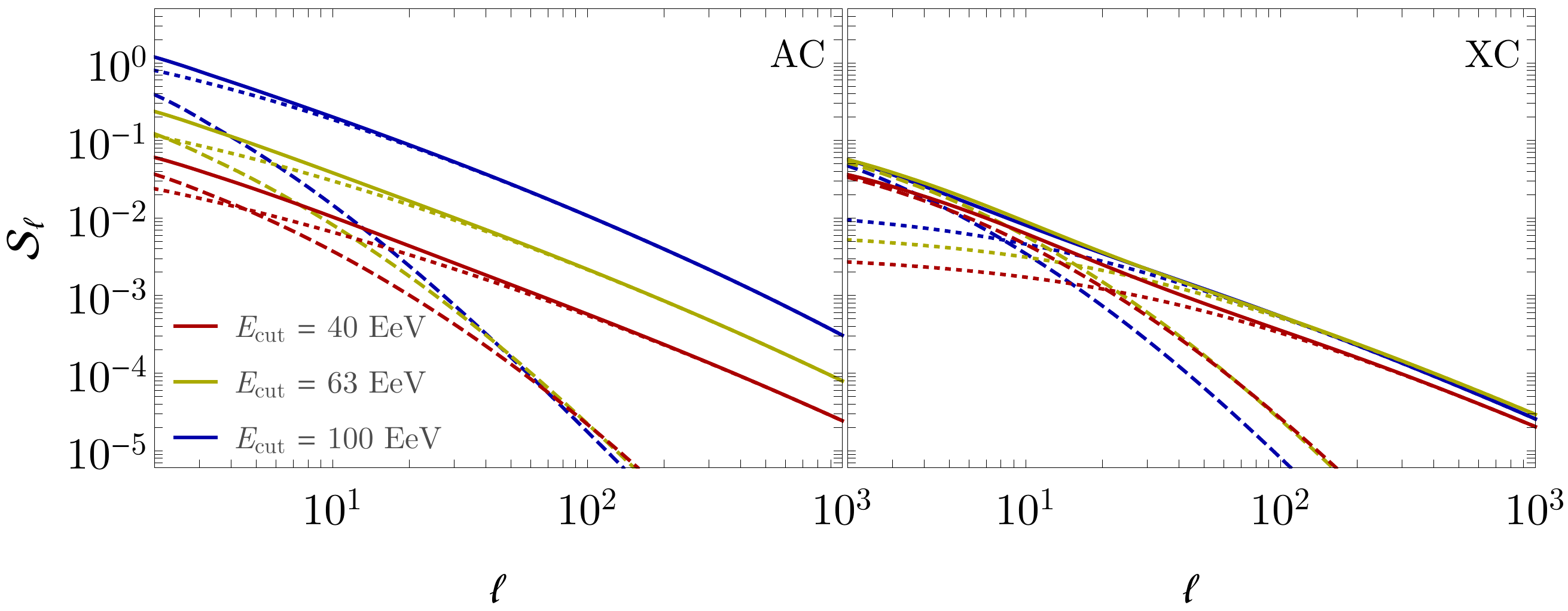}
	\caption{Angular AC and XC power spectra considered in this work. Dotted and dashed curves respectively refer to the 1- and 2-halo contribution to the total signal (solid curves).}
	\label{fig:cl_nl}
\end{figure}

In Fig.~\ref{fig:cl_nl}, we show the expected signal for the AC (left panel) and the XC (right panel).  Colours refer to the three energy cuts discussed above, namely red for \(\Ecut\simeq40\,\mathrm{EeV}\), yellow for \(\Ecut\simeq63\,\mathrm{EeV}\), and blue for \(\Ecut=100\,\mathrm{EeV}\).  The dashed and dotted curves show the 1-halo and 2-halo contributions to the total power spectrum, with the sum of both shown by the solid curves.  For simplicity, we have not included any beam smoothing in the plot.  We can see how the signal for the XC is lower than the AC, as is expected from the fact that the XC mixes two different radial kernels.  If we employ optimal weights for the XC the signal would become identical to that of the AC.  In our simplistic linear treatment of perturbations, this happens because the UHECR and galaxy kernels would be identical. The statistical uncertainties for both correlation functions, however, would be different, given their different shot-noise levels.

\begin{figure}
\centering
    \includegraphics[width=\textwidth]{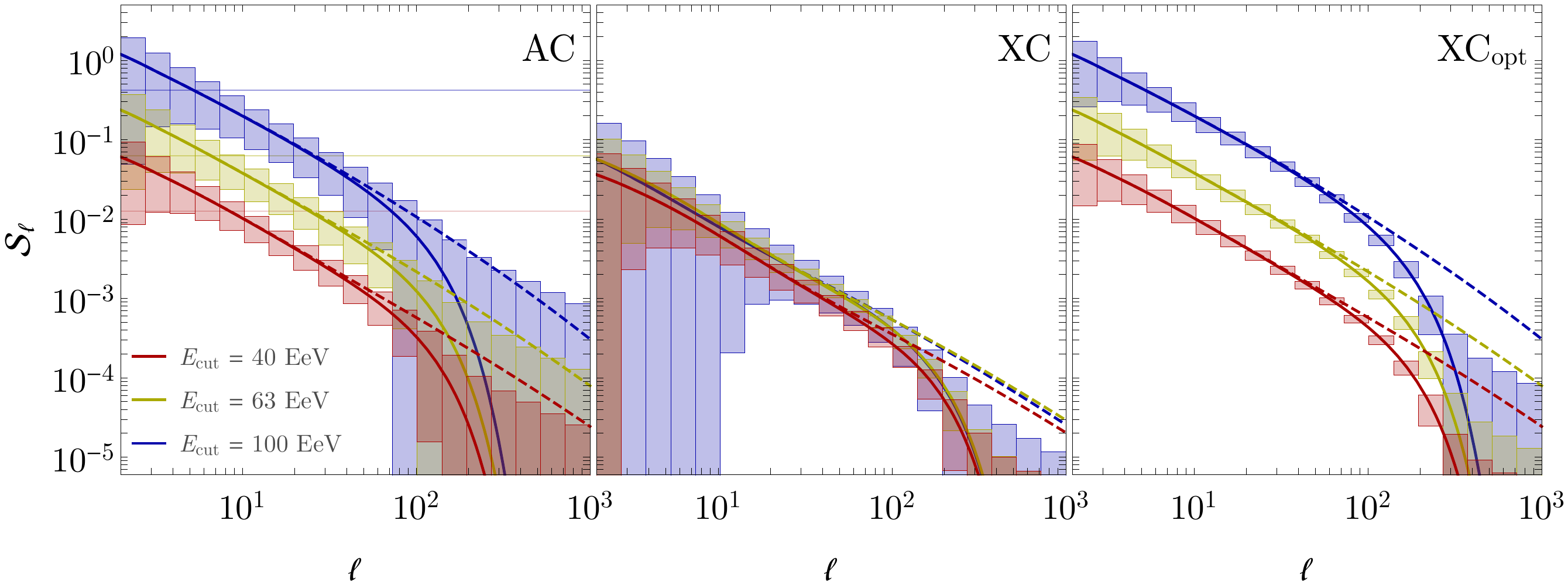}
	\caption{Expected power spectra and \(\ell\)-binned 1\(\sigma\) uncertainties (shaded boxes) including a \(1^\circ\) Gaussian smoothing beam to account for the angular resolution of UHECR experiments (solid curves). For reference, horizontal lines in the leftmost plots denote shot noise levels and the dashed curves show the beam-free prediction.}
	\label{fig:cl_el}
\end{figure}

To understand better the role of the different uncertainties on the theoretical signal, in Fig.~\ref{fig:cl_el} we show again the expected signal as in Fig.~\ref{fig:cl_nl} (solid curves, same colour code) and include a \(1^\circ\) Gaussian smoothing beam to account for the angular resolution of UHECR experiments (for reference, we also show the beam-free prediction as dashed lines). On top of it, we present the corresponding \(\ell\)-binned 1\(\sigma\) error bars as shaded boxes for 20 log-spaces multipole bins between \(\ell_{\rm min}=2\) and \(\ell_{\rm max}=1000\). If we compare the leftmost and central panels, namely AC vs XC, it is easy to see how the range of multipoles where error bars are small enough to allow a detection is larger for XC than for AC for the \(\Ecut\simeq40\,\mathrm{EeV}\) and \(\Ecut\simeq63\,\mathrm{EeV}\) cases.  However, for the sparser UHECR sample with \(E_{\rm cut}=100\,\mathrm{EeV}\) the opposite applies; more precisely, the detectable range of multipoles for the XC is smaller and pushed towards higher \(\ell\) compared to the AC.  This is due to a combination of two factors: for the higher end of UHECR energies the propagation horizon of UHECRs is small, and the UHECR sky looks more anisotropic, boosting the AC.  At the same time, the mismatch in kernels is prominent, the more so the higher the energy, and it drives the XC signal down. Combined with the larger shot noise in the UHECR data, this can explain the performance of the \(100\,\mathrm{EeV}\) case -- indeed, the UHECR shot noise is the main factor that prevents a detection of the signal at mid-\(\ell\) values (the per-\(\ell\) signal is 1\(\sigma\) compatible with zero, see below).

In the rightmost panel of Fig.~\ref{fig:cl_el} we show the XC signal when we apply theoretical optimal weights.  In this case the highest energy set performs the best, and this is expected from the previous arguments: the signal is boosted back up to the same level of the AC because the kernels of galaxies and UHECRs now coincide. Additionally, while the uncertainty increases with energy as both samples become sparser, it is not large enough to hide the XC signal.  It is worth noticing that the increase in galaxy power that we expect towards lower redshifts, is significantly less relevant than the matching of the radial kernels.

In practice, using optimal weights may not be possible given the uncertainties in the radial kernel for UHECRs (we do not know yet the actual injection spectrum). The availability of redshift information in the galaxy catalog, however, would allow us to turn this into an advantage: the UHECR kernel could be reconstructed by modifying the galaxy weights to maximize the signal-to-noise, essentially following the `clustering redshifts' method used to reconstruct unknown redshift distributions in weak lensing data \cite{Newman:2008mb}.

\begin{figure}
\centering
    \includegraphics[width=\textwidth]{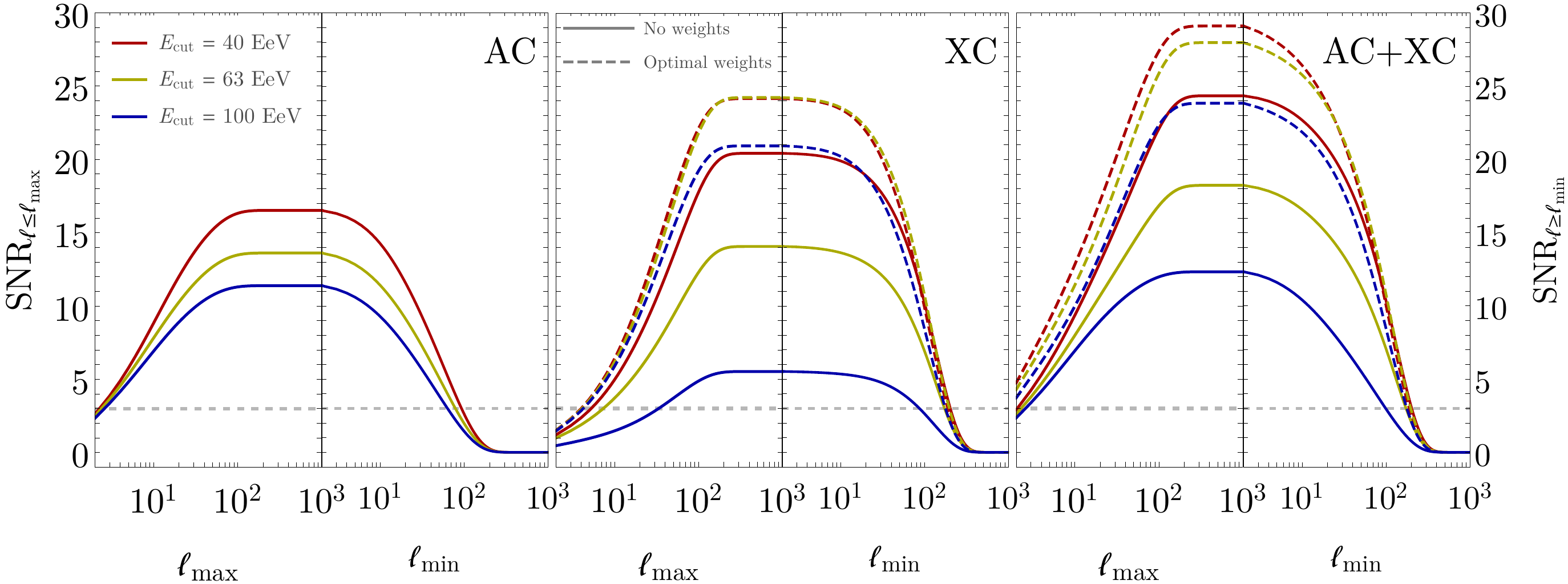}
    \caption{SNR for UHECR flux anisotropies from different combinations of data, namely UHECR AC in the leftmost panel, XC in the central panel, and the combination of all data in the rightmost panel. In each panel, the left half shows the cumulative SNR as a function of the maximum multipole, \(\ell_{\rm max}\), whereas the right half is for the cumulative SNR as a function of the minimum multipole, \(\ell_{\rm min}\). The horizontal dashed line mark the \(3\sigma\) threshold for detection.}
    \label{fig:sn}
\end{figure}

To quantify the improvement in detectability brought by the XC, in Fig.~\ref{fig:sn} we present the cumulative SNR for all the data combinations discussed in Sect.~\ref{ssec:results.fisher}, viz.\ AC alone (leftmost panel), XC alone (central panel), and all the data combined in a single data vector \(\bm S_\ell\) (rightmost panel). In each panel, the left half shows the cumulative SNR as a function of the maximum multipole, \(\ell_{\rm max}\), whilst the right half is for the cumulative SNR as a function of the minimum multipole, \(\ell_{\rm min}\). In both cases, the case with all the data combined has unsurprisingly the largest SNR, but the contributions from AC and XC come from different angular scales, which in turn are sensitive to different redshift ranges, depending upon \(\Ecut\), which sets the propagation depth for UHECRs.  This highlights the complementarity of the two observables.
\begin{figure}
\centering
    \includegraphics[width=\textwidth]{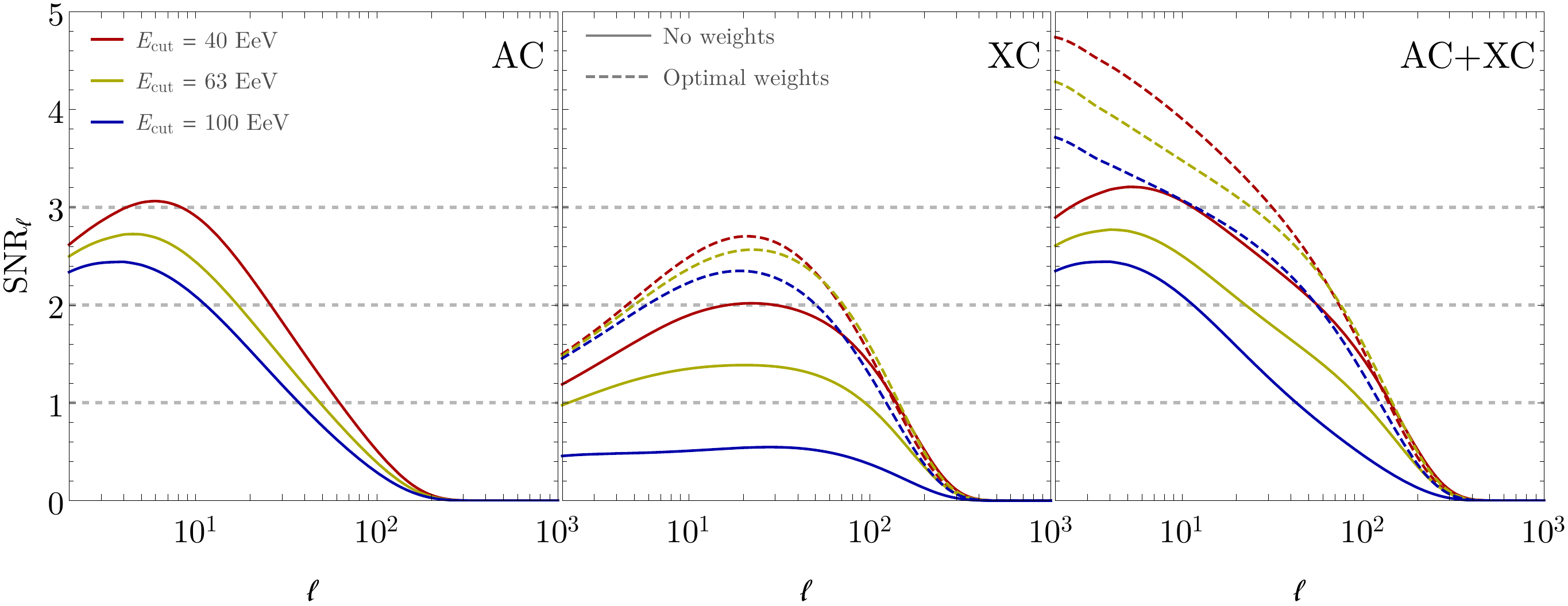}
	\caption{SNR per multipole, \({\rm SNR}_\ell\), for the AC signal, the XC signal with both normal and optimal weights, and their combination AC+XC (leftmost, central, and rightmost panel, respectively). Different colours refer to different energy cuts, and the three horizontal, dashed lines show the thresholds for \(1,\,2\), and \(3\sigma\) detection.}
	\label{fig:sn_el}
\end{figure}

The aforementioned sensitivity to different angular scales can be captured better by looking at Fig.~\ref{fig:sn_el}, where we show the contribution to the total SNR from each integer multipole, \({\rm SNR}_\ell\). The colour code is the same as throughout the paper, and we mark with horizontal dashed lines the thresholds corresponding to \(1,\,2\) and \(3\sigma\) evidence for a one-parameter amplitude fit. These panels can be interpreted as the evidence for anisotropy on a given scale, for which it is clear that the XC with galaxies helps to push the detectability of the signal to smaller scales, i.e., larger \(\ell\) values.  This per-\(\ell\) \(\mathrm{SNR}_\ell\) is a useful quantity to assess whether the AC or the XC is the best observable to detect the anisotropy in UHECRs, assuming that UHECRs trace the LSS.

The sensitivity of the XC to small-scale anisotropies can be precious in realistic situations for two further reasons.  First, a single Earth-based experiment is blind to a large fraction of the sky (roughly speaking one celestial hemisphere); galaxy catalogues can also have incomplete sky coverage.  Moreover, it might be advantageous, see our discussion of the direction-dependent magnetic deflections in Sec.~\ref{ssec:model.MFs}, to restrict the UHECR data set to a portion of the sky to maximise the chances for a clean detection.  In all these situations the low harmonic multipoles are the most affected by these sky cuts.  Second, if two experiments join their data set as the Telescope Array and Pierre Auger collaborations have done in their harmonic AC analysis, they need to cross-calibrate their sets, and this cross-calibration introduces errors that are significantly larger for low multipoles than for high multipoles \cite{Array:2013dra,Aab:2014ila,Deligny:icrc2015,diMatteo:2020dlo}.

\begin{figure}
\centering
    \includegraphics[width=\textwidth]{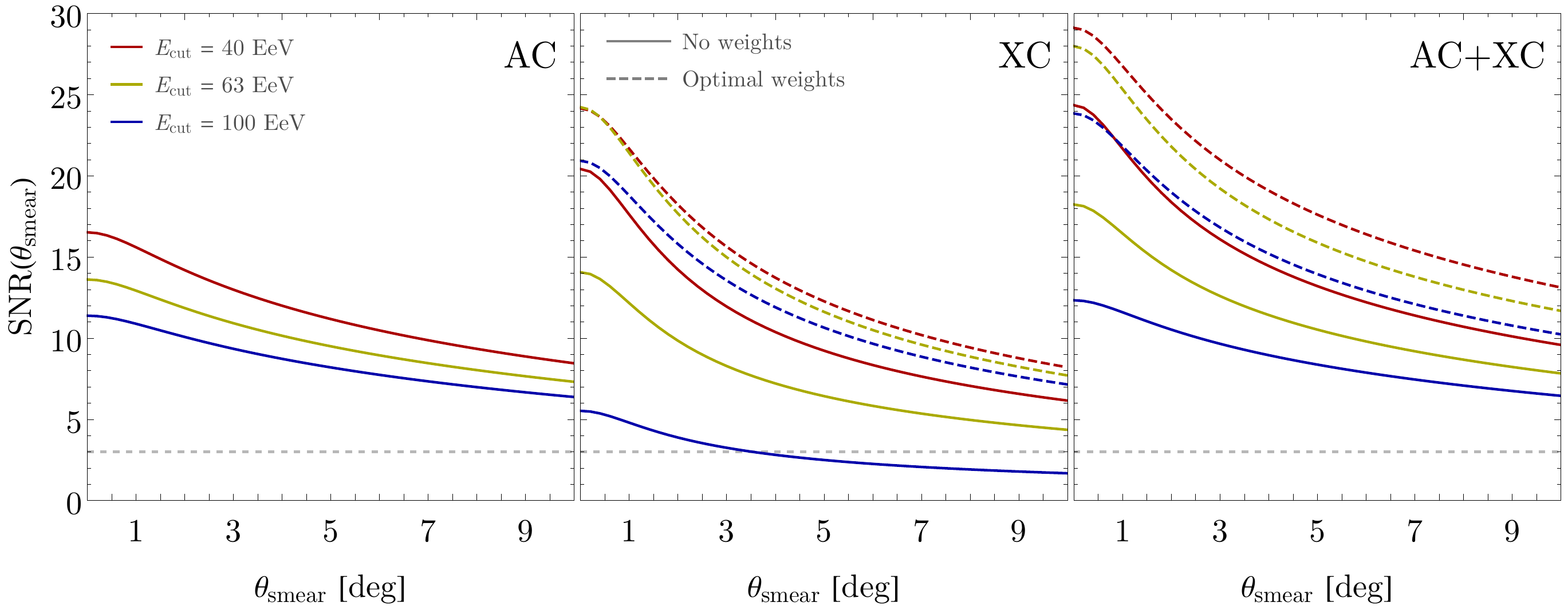}
	\caption{Total SNR, \(\sqrt{\sum_{\ell=\ell_{\rm min}}^{\ell_{\rm max}}{\rm SNR}_\ell^2}\), as a function of the smearing angle for the AC signal, the XC signal with both normal and optimal weights, and their combination AC+XC (leftmost, central, and rightmost panel, respectively). Different colours refer to different energy cuts, and the horizontal, dashed line shows the thresholds for \(3\sigma\) detection.}
	\label{fig:sn_smear}
\end{figure}

As we have argued in Section~\ref{ssec:model.MFs}, there is no shortcut to account for the effects of the GMF on the AC and XC.  Nonetheless, it is instructive to look at how the signal degrades with a simple smearing of the source map.  To this end, we have plotted the total SNR for \(\ell=[2,1000]\) as a function of the smearing angle \(\theta_\mathrm{smear}\) of the galaxy map, for the same energy cuts we have used so far in Fig.~\ref{fig:sn_smear}.  According to Eq.~(\ref{eq:beam}) the deflections for \(40\,\mathrm{EV}\) rigidity peak at around \(7^\circ\) near the Galactic centre, whereas more than half of the sky would be well described by a \(2.5^\circ\) smearing---note that, as we have mentioned in the introduction, the composition of UHECRs at the highest energies is not known \cite{Castellina:2019huz,Bergman:2019aaa}, a heavier composition would imply lower rigidities and larger deflections.  The smearing impacts the high-multipole regions in the XC more than it does for the AC, as expected, and degrades the XC more prominently at larger smearing angles (see also Appendix~\ref{app:gmf}).

The XC is strongly dependent on energy and choice of weights.  This means that we can disentangle small-scale anisotropies caused by the propagation through the GMF from the intrinsic anisotropies inherited from the LSS.  In particular, any GMF-induced signal is erased at higher energy because UHECRs go more straight, whereas the LSS-inherited signal is enhanced because of the smaller propagation horizon.  Moreover, any GMF-induced signal is indifferent to the weights we apply, whereas the signal from LSS anisotropies is strongly enhanced with the use of optimal weights.

Before closing this section, let us remark that in a real experiment there will be modelled and unmodelled systematic errors to take into account.  Systematic errors are expected to contribute to the AC more significantly than to the XC, particularly on large scales (low-\(\ell\) end), e.g., the cross-calibration of two UHECR data sets.  On the other hand, biassed redshift information in galaxy catalogues or UHECR injection properties will affect both observables.  To be clear, whereas the galaxy catalogue and the optimal weights, which depend on UHECR data for the reconstruction of the injection properties, do not enter in the AC obtained from UHECR data alone, they are needed once we test the source model, e.g., that UHECRs correlate with the LSS.  Hence, once systematic effects are taken into account, the SNR for the AC may decrease more than that of the XC.  This is one further motivation to explore the possibilities and improvements from the use of cross-correlations in UHECR anisotropy studies.

% --  --  --  --  --  --  --  --  --  --  --  --  --  --  --  --  --  --  --  --  --  --  --  --  --  -- -
\section{Conclusions and outlook}\label{sec:conclusions}
% --  --  --  --  --  --  --  --  --  --  --  --  --  --  --  --  --  --  --  --  --  --  --  --  --  -- -

In this work, we have introduced a new observable for UHECR physics: the harmonic-space cross-correlation between the arrival directions of UHECRs and the distribution of the cosmic LSS as mapped by galaxies, Eq.~\eqref{eq:cl_limber}.  We have developed the main theoretical tools that are necessary to model the signal and its uncertainties.

The take-away points of this study are:
\begin{itemize}
    \item The cross-correlation can be easier to detect than the UHECR auto-correlation for a range of energies and multipoles (see Figs.~\ref{fig:cl_el} and~\ref{fig:sn_el}).  This performance is mostly driven by the sheer number of galaxies that can trace the underlying LSS distribution, which is assumed to be the baseline distribution for both the UHECR flux and the galaxy angular distribution.
    \item The cross-correlation is more sensitive to small-scale angular anisotropies than the auto-correlation, and vice versa.  It can, therefore, be instrumental in understanding properties of UHECR sources that would not be accessible otherwise.
    \item It is in principle possible to optimize the cross-correlation signal by assigning optimal redshift-dependent weights to sources in the galaxy catalog, to match the UHECR radial kernel as determined by UHECR energy losses.  Since matching the kernels has a strong impact on the cross-correlation, it could be possible to use this effect to reverse-engineer the injection model (which defines the radial kernel).
    \item The great disruptor of UHECR anisotropies is the GMF.  The cross-correlation, with its higher signal-to-noise ratio and sensitivity to small angular scales, could be very useful in understanding the properties of the GMF (although we have not explored this angle here).  Moreover, it may be possible, in the near future, to exploit a tomographic approach to disentangle the effects of intervening magnetic fields from different injection spectra, and study different regions of the sky separately.
\end{itemize}

In our treatment, we do not take any experimental uncertainties into account, besides the experimental UHECR angular resolution. Moreover, we limit ourselves to a proton-only injection model and do not include the effects of the intervening magnetic fields.  This choice was made in order to underline the physics behind our proposal and method, and can be readily generalized and extended to include the (theoretical and experimental) properties of the different galaxy and UHECR catalogs, different injection models, and to separate the number of events and energy cut, in order to best forecast the possibilities of present and upcoming UHECR data sets.

Moreover, in this first work we have made the case for the XC between UHECRs and galaxies, but the logic and methods we have developed can be applied to other XCs with different matter tracers and different messengers. The distribution of visible matter in the sky can be traced not only by galaxies, but also by the thermal Sunyaev-Zeldovich effect. The latter is produced by the inverse Compton scattering of CMB photons by hot electrons along the line-of-sight.  Because a thermal Sunyaev-Zeldovich map is a map of CMB photons, it is very accurate down to angles much smaller than a degree, and its signal peaks at low redshifts \cite{Erler:2017dok}.  This cross-correlation could therefore be useful in further disentangling the astrophysical properties of UHECR sources.

Charged UHECRs are not the only high-energy messengers whose production mechanisms and sources are not known.  Recently, the IceCube collaboration has detected a few high energy astrophysical neutrinos, with energies above a PeV \cite{Aartsen:2014gkd}.  Such neutrinos are expected to be produced in the same extreme astrophysical sources as UHECRs and/or in their immediate surroundings.  The cross-correlations between neutrinos and the LSS will then inform about the properties of the highest-energy astrophysical engines, see~\cite{Fang:2020rvq} (see also \cite{Ando:2015bva}).  Without the use of cross-correlations, because of the very small number of neutrino events in present data, and in the foreseeable future \cite{Fang:2020rvq,Sapienza:2020rte,Allison:2015eky,Nelles_2019}, the detection of the anisotropic pattern could be challenging. Since neutrinos interact extremely weakly, they can propagate unhampered for long distances: their horizon is almost the entire visible Universe.  Therefore, in addition to galaxies, complementary information could be extracted from cross-correlating neutrinos with other tracers, including CMB lensing \cite{Aghanim:2018oex} and cosmic shear surveys \cite{Mandelbaum:2017jpr}, both of which trace the overall matter distribution in the Universe, including both its dark and luminous components, out to higher redshifts with broader kernels (see Refs.~\cite{Fornengo:2014cya,Cuoco:2015rfa,Branchini:2016glc,Ammazzalorso:2019wyr} for the analogous analysis with \(\gamma\) rays).  Measuring these cross-correlations could reveal whether the most energetic particle accelerators in the Universe preferentially reside in high-density visible or dark environments.

\acknowledgments
FU wishes to thank A.\ di Matteo for useful correspondence, and P.~Tinyakov for valuable comments on the manuscript.  FU is supported by the European Regional Development Fund (ESIF/ERDF) and the Czech Ministry of Education, Youth and Sports (MEYS) through Project CoGraDS - \verb|CZ.02.1.01/0.0/0.0/15_003/0000437|. SC is supported by the Italian Ministry of Education, University and Research (\textsc{miur}) through Rita Levi Montalcini project `\textsc{prometheus} -- Probing and Relating Observables with Multi-wavelength Experiments To Help Enlightening the Universe's Structure', and by the `Departments of Excellence 2018-2022' Grant awarded by \textsc{miur} (L.\ 232/2016). DA acknowledges support from the Beecroft Trust, and from the Science and Technology Facilities Council through an Ernest Rutherford Fellowship, grant reference ST/P004474/1. We would like to acknowledge SARS-Cov-2 for the peace of spirit our quarantines in three different countries have given us to finish this work.

% --  --  --  --  --  --  --  --  --  --  --  --  --  --  --  --  --  --  --  --  --  --  --  --  --  -- -
\appendix
% --  --  --  --  --  --  --  --  --  --  --  --  --  --  --  --  --  --  --  --  --  --  --  --  --  -- -
\counterwithin{figure}{section}

% --  --  --  --  --  --  --  --  --  --  --  --  --  --  --  --  --  --  --  --  --  --  --  --  --  -- -
\section{Power spectra}\label{app:pk_cl}

Three-dimensional fields \(\delta_a(\xv)\) can be decomposed into their Fourier modes
\begin{equation}
	\delta_a(\kv)\deq\int \de k^3\;\delta_a(\xv)\,e^{-i\kv\cdot\xv} \,,
\end{equation}
whose covariance is the power spectrum \(P_{ab}(k)\). Assuming statistical homogeneity and isotropy, it is implicitly defined by
\begin{equation}
	\left\langle \delta_a(\kv)\delta_b^\ast(\kv')\right\rangle\deq\delta(\kv-\kv')\,P_{ab}(k) \,,
\end{equation}
where the angle brackets denote averaging over ensemble realizations of the random fields inside them.

Equivalently, two-dimensional fields \(\Delta_a(\nv)\) can be decomposed into their harmonic coefficients
\begin{equation}
	\Delta_{\ell m}^a\deq\int \de \Omega\;Y^*_{\ell m}(\nv)\,\Delta_a(\nv) \,,
\end{equation}
where \(\Omega=(\theta,\varphi)\) is the solid angle on the sky, \(Y_{\ell m}\) are the spherical harmonic functions, \(\hat{\bm n}\) is the line-of-sight direction, and in this work \(a=\left\{{\rm CR},\,g\right\}\). The covariance of the \(\Delta_{\ell m}\) is the angular power spectrum \(\cS^{ab}_\ell\), defined as
\begin{equation}
	\langle \Delta^a_{\ell m}\,\Delta^{b\ast}_{\ell'm'}\rangle\deq\delta_{\ell\ell'}\delta_{mm'}\cS^{ab}_\ell \,.
\end{equation}

For two projected fields, \(\Delta_a\) and \(\Delta_b\), associated to three-dimensional fields \(\delta_a\) and \(\delta_b\) via radial kernels \(\phi_a\) and \(\phi_b\) (as in Eqs~\ref{eq:delta_cr} and \ref{eq:delta_g}), their Fourier- and harmonic-space power spectra are related through
\begin{equation}
	\cS^{ab}_\ell=\frac{2}{\pi}\int \de k\;k^2\,\int \de\chi_1\; \phi_a(\chi_1)\,j_\ell(k\chi_1)\int \de\chi_2\;\phi_b(\chi_2)\,j_\ell(k\chi_2)\,P_{ab}(k;z_1,z_2) \,,
\end{equation}
where \(j_\ell\) is the spherical Bessel function of order \(\ell\). For broad kernels, we can use the Limber approximation, \(j_\ell(x)\sim\sqrt{\pi/(2\ell+1)}\delta(\ell+1/2-x)\), in which case the previous relation simplifies to Eq.~\eqref{eq:cl_limber}.

\section{Optimal weights}\label{app:optweight}
Here we derive the choice of optimal weights discussed in Section \ref{sssec:model.cls.weight}. The derivation is a standard result in statistics and follows the discussion in Appendix A of \cite{Alonso:2020mva}.

Consider a vector of \(N\) measurements \({\bm x}= (x_1,...,x_N)\) and the problem of finding the linear combination of this vector that provides the best estimator of a given quantity \(y\). Assuming Gaussian statistics, the conditional probability is
\begin{equation}
  \log p(y|{\bm x})= {\bm z}^{\sf T}\textsf{\textbf{C}}^{-1}_{zz}{\bm z}-{\bm x}^{\sf T}\textsf{\textbf{C}}_{xx}^{-1}{\bm x},
\end{equation}
where \({\bm z}\equiv (y,x_1,...,x_N)\), and \(\textsf{\textbf{C}}_{ab}\) is the covariance matrix between \({\bm a}\) and \({\bm b}\)\footnote{We use boldface sans serif characters \(\textsf{\textbf{C}}\) to denote matrices, and boldface roman characters \({\bm C}\) to denote vector quantities.}. The minimum-variance estimator for \(y\) given this distribution coincides with its mean, which is given by
\begin{equation}
  \hat{y}={\bm w}^{\sf T}{\bm x}\equiv {\bm C}_{xy}^{\sf T}\textsf{\textbf{C}}_{xx}^{-1}{\bm x}.
\end{equation}
The linear coefficients \({\bm w}\) are the so-called Wiener filter. 

If \(y\) is the UHECR flux and \({\bm x}\) is a set of galaxy overdensity maps at different radial shells with comoving width \(\de\chi\), in the shot-noise dominated regime \({\bm C}_{xy}\) and \(\textsf{\textbf{C}}_{xx}\) are given by
\begin{align}
  &{\bm C}_{{\rm CR}, {\rm g}}(\chi)\propto \de\chi\,\frac{\alpha[z(\chi)]}{1+z(\chi)},\\
  &\textsf{\textbf{C}}_{{\rm g},{\rm g}}(\chi,\chi')\propto\delta_{\chi,\chi'}\de\chi\,\chi^2\bar{n}_{{\rm g,c}}(\chi),
\end{align}
where we have ignored all \(r\)-independent prefactors. Therefore, the Wiener filter in this case is
\begin{equation}
  w(\chi)=\frac{\alpha[z(\chi)]}{[1+z(\chi)]\chi^2\bar{n}_{{\rm g,c}}(\chi)}.
\end{equation}

\section{Halo occupation distribution}\label{app:hod}
Halo occupation distribution models have been used profusely in the literature to describe the connection between the galaxy number density and the matter overdensity. We describe briefly the specifics of the model used here to describe the low-redshift 2MRS sample, which follows \cite{Ando:2017wff}. 

Within the halo model \cite{Peacock:2000qk,Cooray:2002dia}, all matter in the Universe can be found in haloes of different masses, and therefore the fluctuations of a given quantity \(x\) can be described in terms of its distribution around haloes as a function of halo mass \(u_x(r, M)\) (also called the halo profile), and the correlated distribution of haloes on large scales. In this formalism, the power spectrum between two quantities \(x\) and \(y\) receives two contributions, coming from the correlations between mass elements belonging to the same halo and mass elements in different haloes (the so called `1-halo' and `2-halo' terms), as \(P_{xy}=P_{xy}^{1{\rm h}}+P_{xy}^{2{\rm h}}\), with
\begin{align}
  P_{xy}^{1{\rm h}}(k)&=\int\de M\;n_{\rm h}(M)\,u_x(k,M) u_y(k,M),\\
  P_{xy}^{2{\rm h}}(k)&=\left[\int \de M\;n_{\rm h}(M)\,b_{\rm h}(M)\,u_x(k,M) \right]\left[\int\de M\;n_{\rm h}(M)\,b_{\rm h}(M)\,u_y(k,M) \right]P_{\rm lin}(k),
\end{align}
where \(n_{\rm h}(M)\) and \(b_{\rm h}(M)\) are respectively the halo mass function and the halo bias, \(P_{\rm lin}(k)\) is the linear matter power spectrum, and \(u_x(k,M)\) is the Fourier transform of the halo profile.

In our case, we want to model the galaxy overdensity \(\delta_{\rm g}\), and therefore we need to specify the halo galaxy density profile. For this, we use the formalism used in \cite{Ando:2017wff}:
\begin{equation}
  u_{\rm g}(k)=\frac{1}{\bar{n}_{\rm g,c}}\left[N_{\rm c}(M)+N_{\rm s}(M)u_{\rm s}(k,M)\right],
\end{equation}
where \(N_{\rm c}(M)\) and \(N_{\rm s}(M)\) are the number of central and satellite galaxies, the latter of which are distributed according to \(u_{\rm s}(r)\). Centrals and satellites are distributed according to Bernoulli and Poisson distributions respectively. Their mean values and the satellite profile are modelled as a function of mass as:
\begin{align}
  &\bar{N}_{\rm c}(M)=\frac{1}{2}\left[1+{\rm erf}\left(\frac{\log M-\log M_{\rm min}}{\sigma_{\log M}}\right)\right]\\
  &\bar{N}_{\rm s}(M)=\left(\frac{M-M_{\rm min}}{M_1}\right)^\alpha\Theta(M-M_{\rm min})\\
  &u_{\rm s}(r,M)=\frac{\Theta(r_{\rm max,g}-r)}{(r/r_{\rm s,g})(r/r_{\rm s,g}+1)^2},
\end{align}
where \(\Theta\) is the Heaviside function. The free parameters of the model are \(M_{\rm min}\), \(M_1\), \(\sigma_{\log M}\), \(r_{\rm max,g}/r_{\rm s}\), \(r_{\rm s,g}/r_{\rm s}\) and \(\alpha\), where \(r_{\rm s}\) is the mass-dependent halo scale radius. We use the best-fit values found by \cite{Ando:2017wff} for these parameters in our calculation.

\section{On magnetic deflections}\label{app:gmf}
\begin{figure}
\centering
    \includegraphics[width=\textwidth]{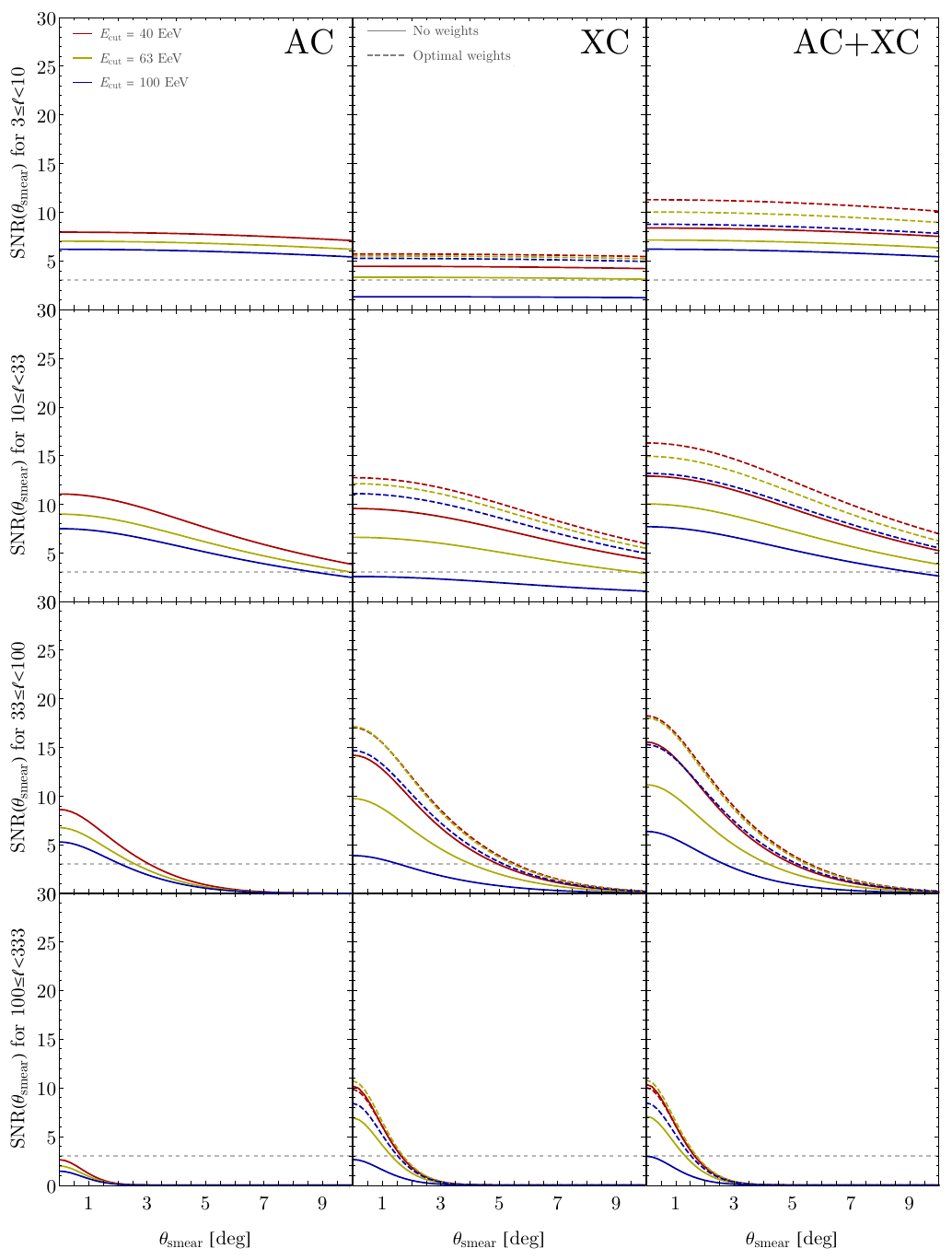}
	\caption{Total SNR per half-decade, \(\sqrt{\sum_{\ell=\ell_{\rm min}}^{\ell_{\rm max}}{\rm SNR}_\ell^2}\), with \((\ell_{\rm min},\ell_{\rm max})\) in \([3,10[\) (top row), \([10,33[\) (second row), \([33,100[\) (third row), \([100,333[\) (bottom row), as a function of the smearing angle for the AC signal, the XC signal with both normal and optimal weights, and their combination AC+XC (leftmost, central, and rightmost panel, respectively, in each row). Different colours refer to different energy cuts, and the horizontal, dashed line shows the thresholds for \(3\sigma\) detection.}
	\label{fig:sn_smear_app}
\end{figure}

The effects of the GMF deflections are not the same across harmonic multipoles. We expect that, for both the AC and the XC, small scales would be more affected by the deflections. We visualise this in Fig.~\ref{fig:sn_smear_app}, where we show the equivalent of Fig.~\ref{fig:sn_smear} but for multipoles in the four half decades: \(\ell\in[3,10[\), \(\ell\in[10,33[\), \(\ell\in[33,100[\), \(\ell\in[100,333[\). As anticipated, the smearing suppresses the power at small scales more incisively, for both the AC and the XC, with the XC being relatively more affected. Nonetheless, in region of the sky where the GMF is small, for example around the Galactic polar cups, the XC has better chances to be detected than the AC at large \(\ell\) (small scales). In reading this figure one should keep in mind at least three simplifications: the larger but physically different effects of the large-scale GMF are not included; the smearing angle is constant across the sky---cf.\ Eq.~(\ref{eq:beam}); the choice of a Gaussian smearing is arbitrary, as other types of beams might reproduce the actual deflections more faithfully.

\bibliography{references}

\end{document}